\documentstyle[a4,12pt]{article}
\input epsf

\newcommand{\Fractal}	{{\rm Fractal}}
\newcommand{\Deficit}	{{\rm Def}}
\newcommand{\Bare}	{{\rm 0}}
\newcommand{\eff}	{{\rm eff}}

\newcommand{\Volume}	{{\rm Volume}}
\newcommand{\Z}		{{\rm Z}}
\newcommand{\QED}	{{\rm QED}}
\newcommand{\QEDf}	{{$\QED_4$}}

\newcommand{\veq}	{
	\setlength{\unitlength}{0.0125in}
	\begin{picture}(12,12)(-6,-4)
	\thicklines
	\put(-1.5, -6){\line( 0, 1){ 12}}
	\put(+1.5, -6){\line( 0, 1){ 12}}
	\end{picture}
}
\newcommand{\vneq}	{
	\setlength{\unitlength}{0.0125in}
	\begin{picture}(12,14)(-6,-4)
	\thicklines
	\put(-5.5, +3){\line( 2,-1){ 12}}
	\put(-1.5, -6){\line( 0, 1){ 12}}
	\put(+1.5, -6){\line( 0, 1){ 12}}
	\end{picture}
}

\newcommand{\Reno}	{{\rm Reno}}

\newcommand{\slashDel}	{\gamma^\mu \partial_\mu}
\newcommand{\pSlash}	{p\hspace{-2.2mm}/}

\newcommand{\pOrtho}	{p_{\bot}}
\newcommand{\Para}	{{/\!/}}
\newcommand{\pPara}	{p_\Para}

\newcommand{\free}	{{\rm free}}
\newcommand{\compact}	{{\rm comp}}

\newcommand{\fig}[1]	{Figure \ref{#1}}

\begin{document}

\baselineskip	= 7mm

\begin{flushright}
\begin{minipage}[b]{33mm} 
 DPNU-96-61\\
 hep-th/9701036\\
 December 1996
\end{minipage}
\end{flushright}

\begin{center}
{\Large\bf
Dynamical Symmetry Breaking\\
in Fractal Space}

\vspace{10mm}

{\large Yukinori Nagatani}
\footnote{E-mail: nagatani@eken.phys.nagoya-u.ac.jp}\\
{\it Department of Physics, Nagoya University, Nagoya 464-01, Japan}

\end{center}

\vspace{10mm}

\begin{center} {\bf ABSTRACT} \end{center}
We formulate field theories in fractal space
and show the phase diagrams
of the coupling versus the fractal dimension
for the dynamical symmetry breaking.
We first consider the $4$-dimensional Gross-Neveu (GN) model
in the $(4-d)$-dimensional randomized Cantor space
where the fermions are restricted to a fractal space by the
high potential barrier of Cantor fractal shape.
By the statistical treatment of this potential,
we obtain an effective action depending on the fractal dimension.
Solving the $1/N$ leading Schwinger-Dyson (SD) equation,
we get the phase diagram of dynamical symmetry breaking
with a critical line
similar to that of the $d$-dimensional $(2<d \leq 4)$ GN model
except for the system-size dependence.
We also consider \QEDf\
with only the fermions formally compactified to $d$ dimensions.
Solving the ladder SD equation,
we obtain the phase diagram of dynamical chiral symmetry breaking
with a linear critical line,
which is consistent with
the known results for $d=4$ (the Maskawa-Nakajima case)
and $d=2$ (the case with the external magnetic field).

\newpage

\section{Introduction}

In nature, we often see the fractals \cite{Mandelbrot}
in the shape of rivers or coastline \cite{CoastLine},
the shape of the electric sparks \cite{Spark},
the crystal growth and so on \cite{Crystal}.
Then we are interested in
what happens in the field theories when a fractal structure
appears as a background structure.

In the quantum gravity
the fractal-shaped space-time often appears \cite{QG},
and we may expect that the quantum field theory in the fractal space-time
can be regarded as a low energy effective theory
of the quantum gravity.

In this paper,
we first give a formulation of
the $4$-dimensional Gross-Neveu (GN) model \cite{NJL}\cite{GN}
in the randomized Cantor space
as a sub-space of the $4$-dimensional space-time.
(See \fig{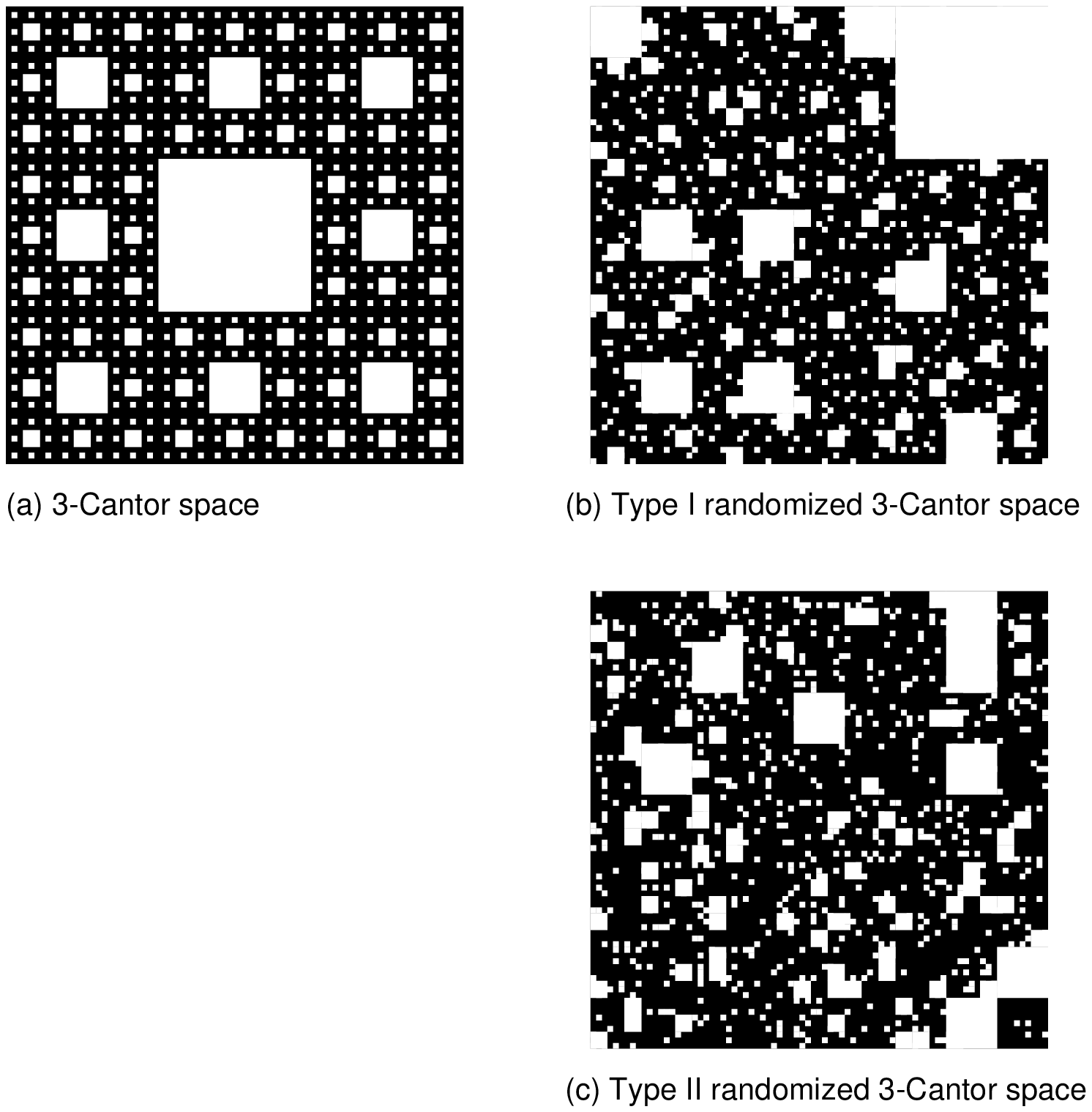}-(b)(c) as an example of
$2$-dimensional randomized Cantor space.)
This randomized Cantor space is a fractal space
which has the fractal dimension $4-d_\Deficit$,
where $d_\Deficit$ means the deficit dimension.
The $(4-d_\Deficit)$-dimensional fractal subspace
has no volume in $4$ dimensions,
then we take fractal renormalization generations $G$ to be large.
We have two formulations of randomization of Cantor space;
the type I randomization and the type II randomization.
In this paper, we use the type II randomization mainly
because of simplicity of the $2$-point function.
We discuss some feature of the type I in Appendix.

In this formulation, we compactify the $4$-dimensional fermion to
the type II randomized Cantor space
by a high potential barrier, namely Cantor potential.
The Cantor space has many holes as space-time deficits,
then we put the high potential barrier
to exclude the fermions from these holes.
We adopt a cut-off level potential as the potential barrier.
In the statistical sense of this randomness
we can define the $n$-point function of the Cantor potentials,
and we show the exact 1 and 2-point functions versus
the deficit dimension $d_\Deficit$ and the fractal generations $G$,
where the 2-point function depends on the space-time configuration $(x,y)$
only through the integer-distance $g(x,y)$.
(The typical 2-point function is described in \fig{2ptFunc2.eps}.)
By means of a shift of the Cantor potential and
the local correlation approximation
of the 2-point function,
this 2-point function generates
the 4-fermion coupling related to the deficit dimensions.

By using the result of the $1/N$ leading Schwinger-Dyson (SD) equation,
we show the phase diagram of the dynamical symmetry breaking
of the GN model in the fractal space
(see \fig{fracPhase.eps}).
This diagram is similar to
that of the GN model in $d$ dimensions \cite{GNlinear},
but the critical line is curved rather than linear.
This phase diagram depends on the system size;
as the system size becomes larger,
the symmetric region in the phase diagram gets smaller.
There exists a critical dimension $4-d_\Deficit^*$ depending on the system size.
In other words, when we fix the fractal dimension,
there exists a critical energy scale
such that the theory in high energy is in the symmetric phase,
while the system in the low energy is in the broken one.

Next we consider a generalization of
the dynamics of fermion 
in the external constant magnetic field in \QEDf\ \cite{Schwinger}
where the dynamical breaking of the chiral symmetry in weak coupling
was shown \cite{Miransky}.
In this theory,
under the background magnetic field the fermion in the lowest
Landau level has no freedom perpendicular to the magnetic field,
and hence it is compactified to the $(1+1)$-dimensional magnetic tube.
This dimensional reduction from $4$ dimensions to $2$ dimensions is
the reason for the chiral symmetry breaking
at any coupling $\alpha > 0$ \cite{Miransky}.

We then give in the last section
a formal generalization of
the $2$-dimensional compactification of charged fermions
by the external magnetic fields \cite{Miransky}
to the $d$-dimensional fermion compactification.
We apply this formulation to the \QEDf,
in which the photon has $4$-dimensional degree of freedom but
the charged fermions have only $d$-dimensional one.

By the ladder SD equation \cite{StQED},
we show the phase diagram of the dynamical chiral symmetry breaking
(See \fig{QEDphase.eps}).
This phase diagram has a linear critical line
which is very similar to
the $d$-dimensional GN model \cite{GNlinear}.
We can understand that
this phase diagram is a kind of analytic continuation between
the $4$-dimensional case without dimensional reduction \cite{StQED} and
the case of the fermion compactified into $2$ dimensions
by the external magnetic field \cite{Miransky}.


\section{Cantor Spaces}

\begin{figure}[b] 
\begin{center}%
 \ \epsfbox{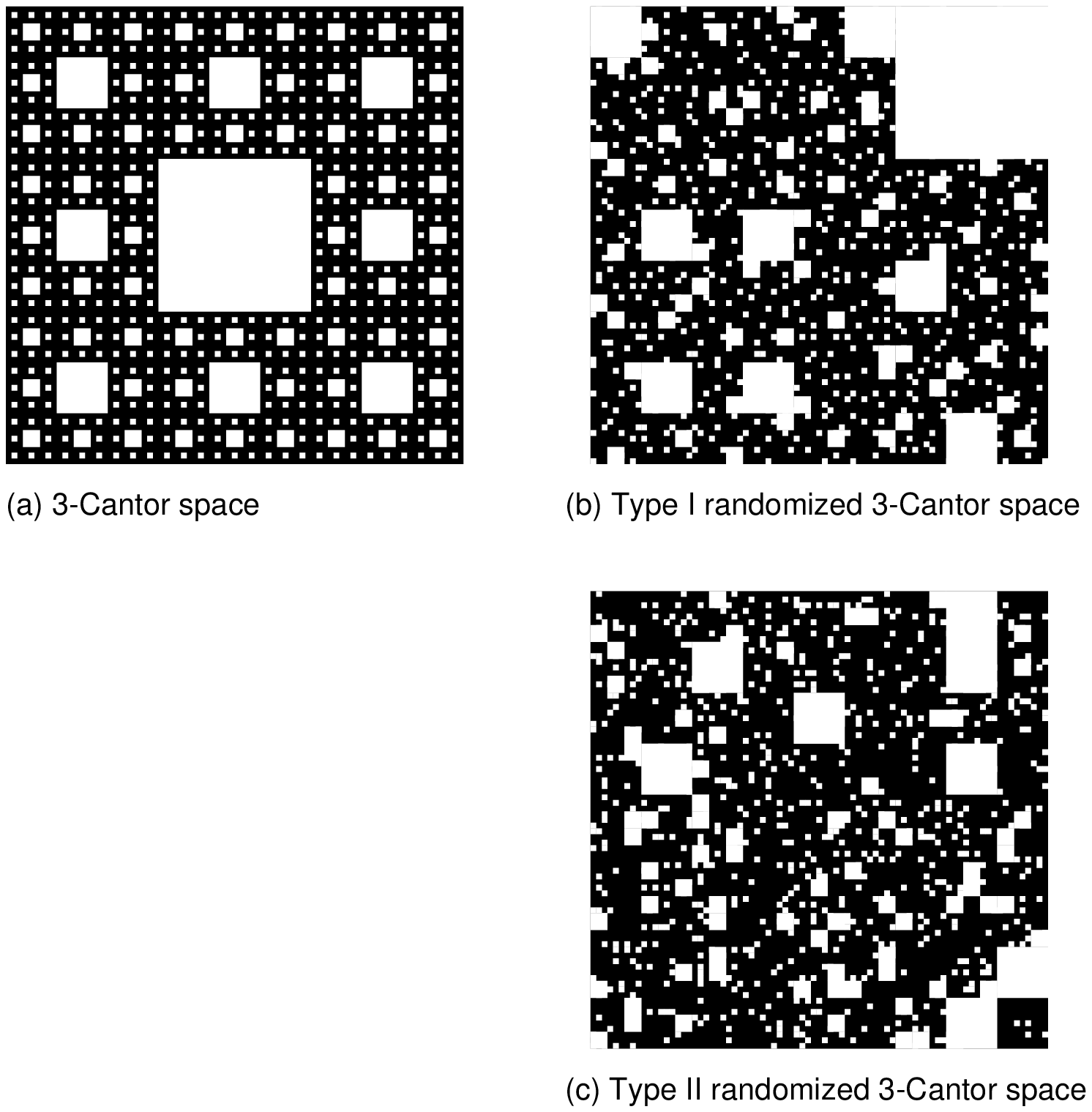}%
 \caption{2-dimensional 3-Cantor spaces.
          {\sf\bf(a)}Ordinary 3-Cantor space.
          Black part means Cantor space and white part means holes. 
          Any square space has a hole at the center position.
          {\sf\bf(b)}Type I randomized 3-Cantor space.
          The difference from (a) is
          only the selection of the hole position out of 9 positions.
          {\sf\bf(c)}Type II randomized 3-Cantor space.
          The number of holes is not constant.
          The average number of holes is equal to (a) and (b).
          Every space has the same fractal dimension
          $d = \log8/\log3 = 1.89\cdots$. }%
          \label{Cantors.eps}%
\end{center}%
\end{figure}

We can make the ordinary $2$-dimensional Cantor Space
by the renormalization sequence as follows (See \fig{Cantors.eps}-(a)).
Suppose that
there is a square space namely ``parent-square''.
At the first step,
we make a $1/3$-scale square hole in the center of this space,
by which there appear $(9-1)$ ``sub-square'' surrounding the center hole.
At the second step,
we zoom up to triple scale,
by which there appear $8$ square spaces,
and we regard them as parent-squares of the next generation.
For any square spaces, we return to the first step.

We call this 2 steps sequence as 1 generation.
The exact Cantor space is generated by infinity-generations sequence.
In any scale, this has a fractal dimension
\begin{eqnarray}
 d_\Fractal &=& \frac{\log \left(3^2-1\right)}{\log 3}
  \;=\; 1.89\cdots, \nonumber
\end{eqnarray}
which is slightly smaller than $2$ dimensions.

Now, we can generalize
the parents-space dimensions from 2 to $D$ and 
the triple scale to $N$-multiple scale $(N,D \in \Z)$,
which we call $D$-dimensional $N$-Cantor space.
This have the fractal dimension
\begin{eqnarray}
 d_\Fractal &=& \frac{\log \left(N^D-1\right)}{\log N}
  \;=\; D - d_\Deficit, \\
 d_\Deficit &\simeq& \frac{1}{N^D \log N},
\end{eqnarray}
where we defined the deficit dimension $d_\Deficit$
which means the effect of holes as the deficit of space.

But this exact D-dimensional $N$-Cantor space has no 
$D$-dimensional volume measure; $\Volume = 0$.
Then we take finite generations $G$ to be large,
and we have a finite volume:
\begin{eqnarray}
 \Volume &=&
  1 - \left[
	    \frac{1}{N^D}
	  + \frac{1}{N^D} \left( \frac{N^D-1}{N^D} \right)
	  + \cdots
          + \frac{1}{N^D} \left( \frac{N^D-1}{N^D} \right)^{G-1}
      \right] \nonumber\\
 &=& \left( 1 - \frac{1}{N^D} \right)^G
 \;\;\simeq\;\; \exp -\frac{G}{N^D}.
\end{eqnarray}

This $N$-Cantor space consists of only one shape,
because there is a square hole in the center of any square space
in any generation.

However,
we can select a position of hole of the same size
at random from $N^D$ positions of sub-space
rather than in the center of parent-square.
Then
we can extend this $N$-Cantor space to randomized ones as a class of spaces
which have the same fractal dimensions (See \fig{Cantors.eps}-(b)(c)).
It is the advantage of this randomized $N$-Cantor spaces class
that we can treat the spaces statistically.

We note that we have two types of randomization.
The type I randomization is to randomize the only position of holes
and fix the number of holes.
The type II randomization is to put holes
with probability $p = 1/N^D$ per $1$ sub-square independently
so that any parent-square has one hole as average.
Each randomization has the same fractal dimension
as the ordinary Cantor space.
We adopt the type II randomization
for simplicity of the $2$-point function.
We will discuss the type I randomized Cantor space in Appendix.

When we consider the field theory in this Cantor space,
we may have two formulations.
The first one is to treat these holes in space as a boundary condition
of fields.
When the space-time dimension $D$ is equal to one,
this formulation is easier
but is very difficult technically
in higher dimensional space like 4 dimensions.
The second one is to treat these holes as high potential barriers
to exclude fermions from these holes.
We put potential
\begin{eqnarray}
 V(x) = \left\{ \begin{array}{ccl}
		 0         & & (x \;\in\; \mbox{ordinary space}), \\
		 \Lambda   & & (x \;\in\; \mbox{holes}),
		\end{array} \right.
\end{eqnarray}
where $\Lambda$ means the cut-off energy of the theory.
We call $V(x)$ the Cantor potential.

For $D$-dimensional randomized $N$-Cantor potential class
which is the potential version of randomized $N$-Cantor space,
we can define the $n$-point function as the statistical meaning of
randomness
\begin{eqnarray}
 \left< V(x) \right>_{N(d)}
    &:=& \int_{N(d):\rm fix} {\cal D}V \; V(x), \nonumber\\
 \left< V(x)V(y) \right>_{N(d)}
    &:=& \int_{N(d):\rm fix} {\cal D}V \; V(x) V(y),\\
    &\vdots& \qquad,
\end{eqnarray}
with normalization $\int_{N(d):\rm fix} {\cal D}V = 1$.

We can calculate the 1-point function easily
as the average of the co-space volume of $N$-Cantor space:
\begin{eqnarray}
 \frac{1}{\Lambda} \left< V(x) \right>_{N(d)} &=&
	    p \:+\: (1-p)p \:+\: \cdots \:+\: (1-p)^{G-1}p
 \nonumber\\
 &=& 1 - \left( 1 - p \right)^G
  =  1 - \left( 1 - \frac{1}{N^D} \right)^G , \label{1ptFunction}
\end{eqnarray}
where we have defined
\begin{eqnarray}
 p &=& \frac{1}{N^D} \label{probability}
\end{eqnarray}
as a probability of the existence of hole per sub-square.

\section{2-Point Function}

To get the 2-point function of the $N$-Cantor space class,
we use renormalization analysis.
Let the $1$-generation be the finest or high-energy one
at the cut-off scale $\Lambda$ 
and the $G$-generation be the system-size one.

We can represent any point $x$
in the $D$-dimensional $G$-generation $N$-Cantor space
as the $G$-columns of $N^D$-digit numbers:
\begin{eqnarray}
&\begin{array}{ccccccccccc}
  x & = & ( & x_G  & x_{G-1} & \cdots & x_i & \cdots & x_2 & x_1 & ) \\
    &   & \multicolumn{9}{c}
           {\mbox{low-resolution} \longleftrightarrow
	    \mbox{high-resolution},}
 \end{array}&
\end{eqnarray}
where $x_i$ has values from $1$ to $N^D$
as a position of sub-square in $i$-generation;
the point $x$ exists in the $x_i$-th sub-square.

In general, if we pick up any two points $x$ and $y$ in the $N$-Cantor space,
we can regard them as the same point
with respect to resolutions from the $G$-generation to
some generation $g+1$ $(0 \leq g \leq G)$,
and as different points with respect to higher resolutions
from $g$-generation to the finest generation $1$:
\begin{eqnarray}
&\begin{array}{ccccccccccc}
  x & = & ( & x_G  & x_{G-1} & \cdots & x_{g+1} & x_g & \cdots & x_1     & ) \\
    &   &   & \veq & \veq    & \cdots & \veq  & \vneq & \cdots & (\vneq) &   \\
  y & = & ( & y_G  & y_{G-1} & \cdots & y_{g+1} & y_g & \cdots & y_1     & ) \\
    &   & \multicolumn{9}{c}
           {\mbox{low-resolution} \longleftrightarrow
	    \mbox{high-resolution},}
 \end{array}& \label{Distance}
\end{eqnarray}
In other words,
for any two points $x$ and $y$,
there exists a certain integer $g = g(x,y) \geq 0$
such that it satisfies (\ref{Distance}).
We can prove that $g(x,y)$ satisfies
\begin{eqnarray}
  g(x,x) &=& 0, \nonumber \\
  g(x,y) &=& g(y,x), \nonumber \\
  g(x,z) &\leq& g(x,y) + g(y,z),
\end{eqnarray}
namely $g(x,y)$ is a distance in the Cantor space.
We call $g(x,y)$ a integer-distance.

When we fix the generations $G$,
for any pairs of points $(x, y)$ such that $g(x,y)$ is a constant,
we cannot distinguish pairs with respect to
the statistical feature of pair.
Then the 2-point function depends on the space-time configuration $(x,y)$
only through the integer-distance $g(x,y)$:
\begin{eqnarray}
 \frac{1}{\Lambda^2} \: \left<\: V(x)V(y) \:\right>_{N,G}
  &=& L\left( g(x,y) ; G \right)_{N}. \label{DefOfL}
\end{eqnarray}
This $L(g;G)$ depends on the parents-space dimension $D$,
integer $N$ which determines the fractal dimension of Cantor space,
the generations of system $G$,
and the integer-distance $g$ as the space-time dependence of 2-point function.
$L(g;G)$ is a decreasing function of $g$.
We call $L(g;G)$ a level, $L(G;G)$ a base level and $L(0;G)$ a top level.
The dependence of each levels on $D$ and $N$ is only through the
probability $p$ in (\ref{probability}).
The top level $L(0;G)$ is self-correlation, and hence
\begin{eqnarray}
 L(0;G)_{N,G}
      &=&  \frac{1}{\Lambda^2} \left<V(x)V(x)\right>
     \;=\; \frac{1}{\Lambda}   \left<V(x)\right> \nonumber\\
      &=& 1 \:-\: \left( 1-p \right)^G.
\end{eqnarray}

By the way, we can also get this result by renormalization transformation:
\begin{eqnarray}
 L(0;1) &=& p, \nonumber\\
 L(0;2) &=& p \;+\; (1-p) L(0;1), \nonumber \\
        &\vdots& \nonumber \\
 L(0;G+1) &=& p \;+\; (1-p) L(0;G) \nonumber \\
        &=& \Reno\left[ L(0;G) \right],
\end{eqnarray}
where we defined
\begin{eqnarray}
 \Reno\left[ X \right] &:=& p \;+\; (1-p) X. \label{Reno}
\end{eqnarray}
This ``Reno'' transformation is
a renormalization transformation of probability
from smaller generation to larger one.
The first term of (\ref{Reno}) is a probability of existence of larger hole
and the second term is that of inheritance of the fine structure.
The result of $n$-times of ``Reno'' transformations is
\begin{eqnarray}
 \Reno^{(n)}\left[ X \right]
  &=& 1 \;-\; (1-p)^n \;+\; (1-p)^n \: X \nonumber \\
  &=& L(0;n) \;+\; \left( 1 - L(0;n) \right)^n X.
\end{eqnarray}

For the base level $L(G;G)$,
the distance between $x$ and $y$ is maximum with respect to $g(x,y)$,
and there is no correlation between potential values $V(x)$ and $V(y)$.
The base level is a probability that
both $V(x)$ and $V(y)$ have non-zero value,
then this level is only the square of these probabilities:
\begin{eqnarray}
 L(G;G) &=& \left( L(0;G) \right)^2.
\end{eqnarray}

\begin{figure}[b] 
\begin{center}%
 \ \epsfbox{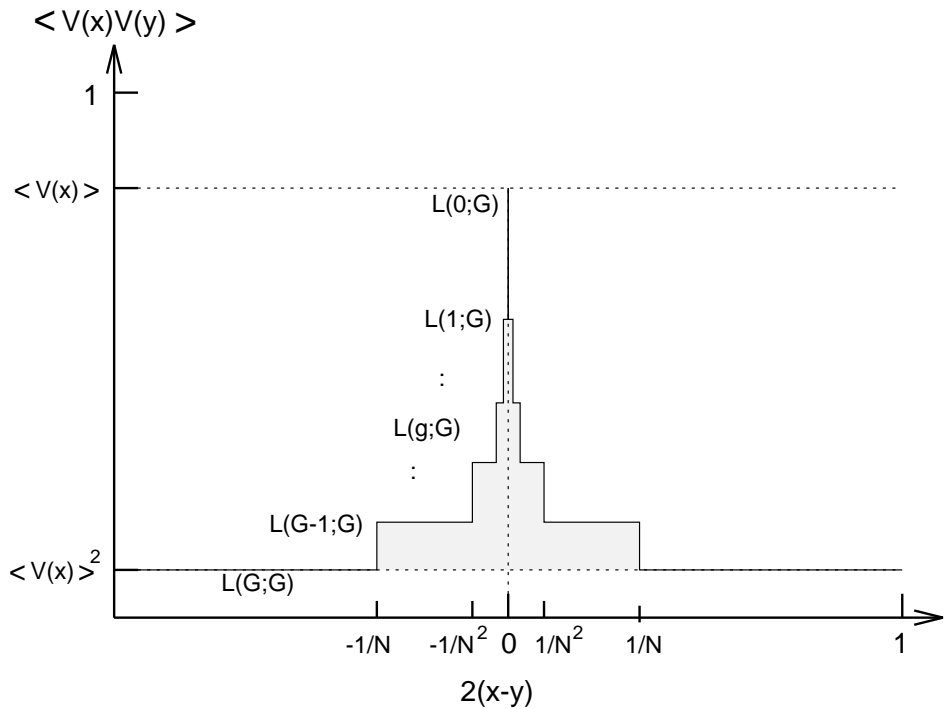}%
 \caption{Typical 2-point correlation function of
	  the 1-dimensional type II randomized $N$-Cantor potential.
	  The exact form depends on both $x$ and $y$
	  through the integer-distance $g(x,y)$.
	  For $D$-dimensional case, we get the 2-point correlation function
	  by replacing $N$ by $N^D$ in $L(G;g)$.
	  In the local potential approximation,
	  we define the weight of delta function $\Omega$
	  as the volume of the shadow zone.}%
	\label{2ptFunc2.eps}%
\end{center}%
\end{figure}

For the intermediate levels $L(g;G) \;\; (0 \leq g \leq G)$,
in the way similar to the top levels we have
\begin{eqnarray}
 L(g;g)     &=& \left( L(0;g) \right)^2, \nonumber\\
 L(g;g+1)   &=& p \;+\; (1-p) \: L(g;g), \nonumber\\
       &\vdots&                          \nonumber\\
 L(g;G+1)   &=& p \;+\; (1-p) \: L(g;G)  \nonumber\\
            &=& \Reno\left[ L(g;G) \right].
\end{eqnarray}
Then the intermediate levels are obtained from the base level
by $\Reno$ transformations:
\begin{eqnarray}
 L(g;G) &=& \Reno^{(G-g)}\left[ L(g;g) \right] \nonumber\\
        &=& L(G-g;G) \;+\; \left( 1 - L(G-g;G) \right)
			 \left( L(g;g) \right)^2  \nonumber\\
        &=& 1 \;-\; 2 (1-p)^G \;+\; (1-p)^{G+g} \nonumber\\
        &=& 1 \;-\; 2 \left( 1 -\frac{1}{N^D} \right)^G
	      \;+\;   \left( 1 -\frac{1}{N^D} \right)^{G+g}.
	    \label{2ptFunction}
\end{eqnarray}
This result is consistent with
both the top level $L(0;G)$ and the base level $L(G;G)$.
We show the typical resultant levels in \fig{2ptFunc2.eps}
when the parent-space dimension $D$ is $1$.

\begin{figure}[b] 
\begin{center}%
 \ \epsfbox{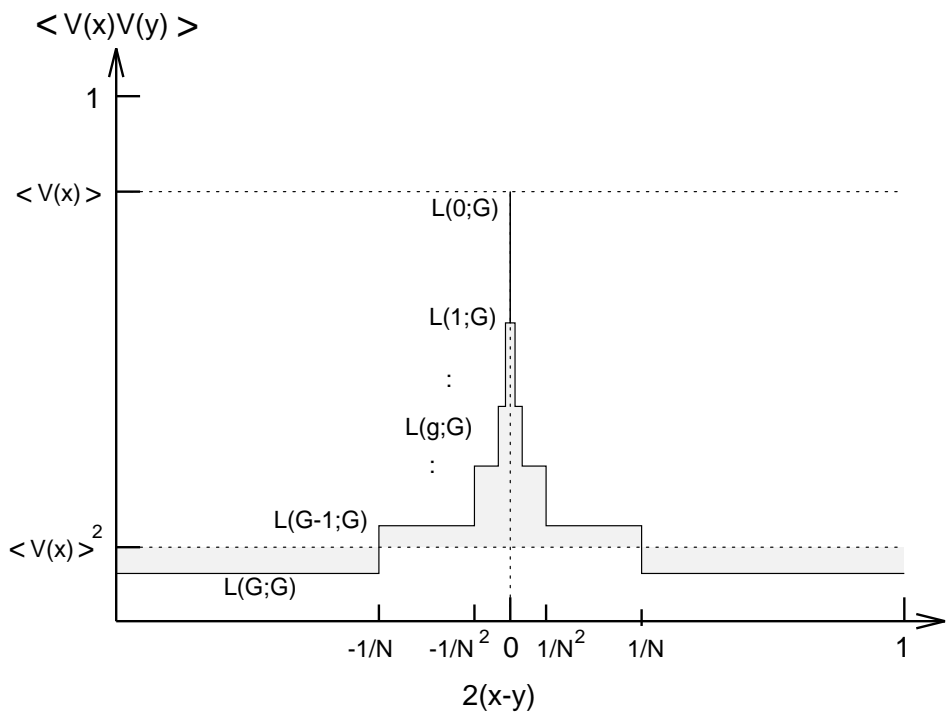}%
 \caption{Type I typical 2-point correlation function of
	  the 1-dimensional randomized $N$-Cantor potential.
	  There are non-local parts. (See Appendix.)}%
	\label{2ptFunc1.eps}%
\end{center}%
\end{figure}

\section{Local Correlation Approximation}

For our purpose,
we shift the potential
\begin{eqnarray}
 V(x) &\longrightarrow& \tilde{V}(x) = V(x) - \left<V(x)\right>
\end{eqnarray}
for $\left<\tilde{V}(x)\right> = 0$.
Then the 2-point functions become
\begin{eqnarray}
 \left<\tilde{V}(x)\tilde{V}(y)\right>
  &=& \left<V(x)V(y)\right> - \left<V(x)\right>^2.
\end{eqnarray}

The base level of $\left<V(x)V(y)\right>$ is equal to
$\left<V(x)\right>^2$,
then $\left<\tilde{V}(x)\tilde{V}(y)\right>$
has no base level,
and has only one peak in $x=y$ point.
Then we approximate this 2-point function
by the Dirac delta-function, namely ``Local Correlation Approximation'':
\begin{eqnarray}
 \left<\tilde{V}(x)\tilde{V}(y)\right>
  &\simeq& \frac{\Lambda^2}{\lambda^D} \:
	    \Omega \: \delta^{(D)}(x-y),   \label{LCA}
\end{eqnarray}
where $\lambda$ has a mass dimension which is a mass scale of the system size,
and $\Omega$ means dimensionless volume of this peak
(See \fig{2ptFunc2.eps}).
$\Omega$ can be estimated through rewriting
2-point function of $\tilde{V}(x)$:
\begin{eqnarray}
\left<\tilde{V}(x)\tilde{V}(y)\right>
  &=& \Delta^G_{g(x,y)} \:+\: \Delta^G_{g(x,y)+1} \:+\: \cdots \:+\:
      \Delta^G_{G-1},
\end{eqnarray}
where
\begin{eqnarray}
 \Delta^G_g
  &:=& L(g;G) \:-\: L(g+1;G) \nonumber\\
  &=&  p \: (1-p)^{G+g}  \nonumber\\
  &=&  \frac{1}{N^D} \left(1-\frac{1}{N^D}\right)^{G+g}
\end{eqnarray}
for $0\leq g < G$.
Then the volume of peak is
\begin{eqnarray}
 \Omega
    &=&  \sum_{g=0}^{G-1} \; \frac{N^{Dg}}{N^{DG}} \: \Delta^G_g \nonumber\\
    &=&  \frac{1}{N^D} \;
         \frac{1}{N^D-2} \;
         \left(1-\frac{1}{N^D}\right)^{2G}
	 \left\{ 1 - \left(\frac{1}{N^D-1}\right)^G \right\} \nonumber\\
   &\simeq& \frac{1}{N^{2D}} \; \left(1-\frac{1}{N^D}\right)^{2G}.
\end{eqnarray}

\section{Gross-Neveu Model in Fractal Space}

The action of the 4-dimensional Gross-Neveu model \cite{GN} is
\begin{eqnarray}
 S[\psi]
  &:=&  \int d^4x \;\; \bar\psi i\slashDel \psi
		\;+\; \frac{G_\Bare}{2N_F} \: (\bar\psi \psi)^2,
\end{eqnarray}
where $G_\Bare$ is the bare coupling constant whose mass dimension is $-2$
and $N_F$ is the number of fermions.
Through the $1/N_F$ expansion analysis it is known that
there is a critical coupling $G^*_\Bare = 4\pi^2/\Lambda^2$.
If $G_\Bare < G^*_\Bare$, then there is no dynamical mass of the fermion and
the theory has an unbroken discrete chiral symmetry,
whereas for $G_\Bare > G^*_\Bare$ the dynamical mass is generated
and the theory is in a dynamically broken phase of the symmetry.

The action of the Gross-Neveu model with extra potential $\tilde{V}(x)$
is given by
\begin{eqnarray}
 S[\psi,V]
  &:=&  S[\psi] \;-\;
        \frac{1}{\sqrt{N_F}} \int d^4x \; V(x) \; \bar\psi \psi,
\end{eqnarray}
where we omitted the tilde from $\tilde{V}$.
We define the effective action of the Gross-Neveu model
with randomized $N$-Cantor potentials
in Euclidean space:
\begin{eqnarray}
 e^{-\tilde{S}_\eff[\psi]}
  &:=& \int {\cal D}V \; e^{-\tilde{S}[\psi,V]},
\end{eqnarray}
where $\tilde{S} = Kinetic + Potential$ means Euclidean action
(when Minkowski action is $S = Kinetic - Potential$).
To get the effective action we approximate this functional integral as:
\begin{eqnarray}
e^{-Potential}
  &=& 1 \:-\: \int d^4x \; \left<V(x)\right> (\bar\psi\psi)(x) \nonumber\\
  & & \quad  \:+\: \frac{1}{2} \: \int d^4x \int d^4y \;
                   \left<V(x)V(y)\right> \;
      (\bar\psi\psi)(x)(\bar\psi\psi)(y) \:+\: \cdots \nonumber\\
  &\simeq&  \exp\left[ \;
		\frac{1}{2} \: \int d^4x \: \int d^4y \;
	        \left<V(x)V(y)\right>
	        \;(\bar\psi\psi)(x)(\bar\psi\psi)(y)
		\; \right],
\end{eqnarray}
where we used the definition of $n$-point function
and absence of 1-point function,
and we neglected more than 3-point functions as an approximation.
Then the approximated effective action is
\begin{eqnarray}
 S_\eff[\psi]
  &:=&  \int d^4x \;\; \bar\psi i\slashDel \psi \nonumber\\
  &+&   \int d^4x \; \int d^4y \;
             \frac{1}{2N_F}
             \left[     \: G_\Bare \: \delta(x-y) \;+\;
	                \left<V(x)V(y)\right> \:
	     \right] \;    \nonumber\\
  & & \qquad \left(\bar\psi\psi\right)(x) \;
	     \left(\bar\psi\psi\right)(y)    \nonumber\\
  &\simeq& \int d^4x \;\; \bar\psi i\slashDel \psi
              \;+\; \frac{G_\eff}{2N_F} \; (\bar\psi\psi)^2
  \label{effectiveAction}
\end{eqnarray}
where we used local correlation approximation in the last equation and
we defined effective $4$-fermion coupling constant
\begin{eqnarray}
 G_\eff &=& G_\Bare \;+\; \frac{\Lambda^2}{\lambda^4} \Omega.
\end{eqnarray}

Our resultant effective action (\ref{effectiveAction})
takes the same form as the one of the original GN model,
and the coupling constant is shifted as the effect of the fractal potential.

Let us define the dimensionless $4$-fermion coupling:
\begin{eqnarray}
 g_\Bare &:=& \frac{\Lambda^2 G_\Bare}{4\pi^2}, \quad \nonumber\\
 g_\eff  &:=& \frac{\Lambda^2 G_\eff }{4\pi^2}, \nonumber
\end{eqnarray}
then the effective coupling reads
\begin{eqnarray}
 g_\eff &=&  g_\Bare
       \;+\; \frac{1}{4\pi^2} \left(\frac{\Lambda}{\lambda}\right)^4 \Omega,
\end{eqnarray}
and the critical coupling in $4$ dimensions becomes $g_\Bare^* = 1$.
The equation of the critical line of the dynamical symmetry breaking becomes
\begin{eqnarray}
 g_\Bare \;+\;
 \frac{1}{4\pi^2} \left(\frac{\Lambda}{\lambda}\right)^4 \Omega &=& 1.
   \label{CriticalLineG0}
\end{eqnarray}

\begin{figure}[b] 
\begin{center}%
 \ \epsfbox{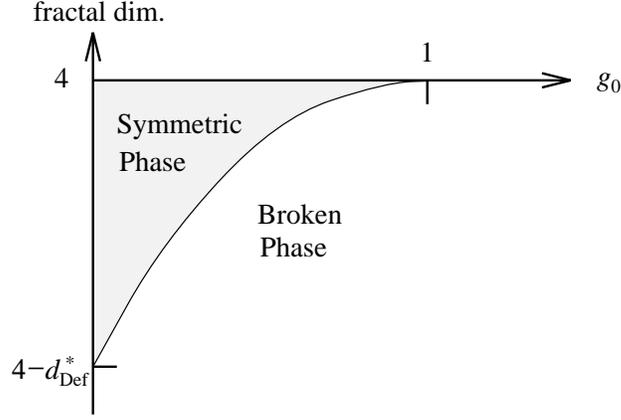}%
 \caption{Phase diagram of the Gross-Neveu model in the $4$-dimensional
	  $N(d)$-Cantor space which has fractal dimension $(4-d_\Deficit)$.
	  In this diagram $g_\Bare$ is the ordinary dimensionless
	  4-fermion bare coupling,
	  and $4-d_\Deficit^*$ means the critical fractal coupling.
	  This phase diagram depends on $\Lambda/\lambda$
	  where $\Lambda$ is the cut-off of $4$-fermion theory and
	  $\lambda$ is the scale of the system size.
	  The critical deficit dimension depends on the system size
	  like
	  $d_\Deficit^* \simeq \frac{2\pi}{c}\left(\lambda/\Lambda\right)^2$.}
          \label{fracPhase.eps}%
\end{center}%
\end{figure}

Now we can discuss the phase diagram
in fractal dimension $4-d_\Deficit$ versus bare coupling $g_\Bare$.
By using
\begin{eqnarray}
 \Omega &=& \frac{1}{N^8} \:
	\left( 1 - \frac{1}{N^4} \right)^{2(G-1)}, \nonumber\\
 d_\Deficit &=& \frac{1}{N^4\:\log N}, \nonumber\\
 \frac{\Lambda}{\lambda} &=& N^G,
\end{eqnarray}
we further rewrite the equation of critical line (\ref{CriticalLineG0}) into
\begin{eqnarray}
   g_\Bare \;+\;
   \frac{d_\Deficit^2 \; (\log N(d_\Deficit))^2}{4\pi^2} \;
   \left(\frac{\Lambda}{\lambda}\right)^{4-2d_\Deficit}
  &=& 1 \label{CriticalLineEq}
\end{eqnarray}
which gives the phase diagram shown in \fig{fracPhase.eps}.
This phase diagram depends on the cut-off scale ratio $\Lambda/\lambda$.
If we fix the ultraviolet cut-off $\Lambda$,
then the phase depends on the system scale $\lambda$.
When system size gets larger,
the symmetric phase region in the phase diagram gets smaller.

There exists a critical dimension $4-d_\Deficit^*$
which depends on the scale ratio $\lambda/\Lambda$.
When we approximate $\log N(d_\Deficit)$ to a constant $c$,
we get a critical deficit dimension:
\begin{eqnarray}
d_\Deficit^* &\simeq& \frac{2\pi}{c}\left(\lambda/\Lambda\right)^2.
\end{eqnarray}
If we fix the fractal dimension,
then there is a kind of critical scale:
\begin{eqnarray}
\lambda^* &\simeq& \Lambda \; \sqrt{\frac{d_\Deficit \; c}{2\pi}},
  \label{CriticalScale}
\end{eqnarray}
above which (in the high energy) the system is in symmetric phase,
while in the low energy it is in the broken phase.



\section{Strong \QEDf\ with $d$-dimensional fermion}

In the \QEDf\ with background constant magnetic field,
it was found \cite{Miransky} that
the dynamical mass of fermion $m_d$ is generated
even in the weak coupling region,
since the 4-dimensional fermion is compactified
to a 2-dimensional one in the lowest Landau level (LLL):
\begin{eqnarray}
 S\left(\pPara,\pOrtho\right)
  &=& i \frac{1}{\pSlash_\Para-m_d} \; e^{-l^2 \pOrtho^2} \; P_{\rm spin} \;,
\end{eqnarray}
where $P_{\rm spin}$ means the spin projection.
This propagator has two directions;
$\pOrtho$ is a direction perpendicular to the constant magnetic field,
while $\pPara$ is the rest which includes the time direction.
For $\pOrtho$ direction, the propagator is Gaussian-damping
with length scale $l = 1 / \sqrt{eB}$ which is the scale of the magnetic field.
We call this direction a compact one.
As for the $\pOrtho$ direction, it is simply a free propagator
with dynamical mass $m_d$ which will be determined
by SD equation later.
We call this direction a free one.

Then we will formally generalize this 2-dimensional
compactification to $d$-dimensional one,
and see whether or not the fermion mass generation takes place.
We assume 4-dimensional momentum $p$ and its volume measure $d^4p$
can be decomposed into two-parts:
\begin{eqnarray}
 p    &=& p_\free + p_\compact, \nonumber \\
 d^4p &=& d^dp_\free \;\; d^{4-d}p_\compact, \nonumber \\
      & & p_\free \; \bot \; p_\compact,
\end{eqnarray}
where $p_\free$ means the $d$-dimensional momentum like $p_\Para$,
and $p_\compact$ means the $(4-d)$-dimensional momentum like $\pOrtho$.
We call $p_\free$ as a free direction
and $p_\compact$ a compactified one.

Based on the analogy with the constant magnetic case,
we assume the fermion propagator as
\begin{eqnarray}
 S\left(p_\free,p_\compact\right)
  &=& i \frac{1}{\pSlash{}_\free-m_d} \;\; e^{-l^2 p_\compact^2} \;,
 \label{CompactPropagator}
\end{eqnarray}
where $l$ means the compactified length like a magnetic scale $1/\sqrt{eB}$.
This propagator is very formal,
but after formally contraction of spinors
we can construct SD equation based on this.
Thus 
our starting point is just the SD equation.
In this propagator there are no spin projections,
but in $d=2$ the SD equation
coincides with
that of the \QEDf\ with background magnetic field \cite{Miransky}.

In the case of the background magnetic field the photon propagator
is the same as the 4-dimensional one without magnetic field.
Then in our case we assume the same Landau-gauge photon propagator:
\begin{eqnarray}
 D_{\mu\nu}(k) &=& \frac{i}{k^2}
  \left[g_{\mu\nu} - \frac{k_\mu k_\nu}{k^2}\right].
\end{eqnarray}

The Euclidean ladder SD equation in 4 dimensions
with the bare mass equal to zero is
\begin{eqnarray}
 m(p^2) &=& \frac{3g^2}{(2\pi)^4}
  \int_{0}^{\Lambda^2} d^4k \;
     \frac{m(k^2)}{k^2 + m^2(k^2)} \;
     \frac{1}{(k-p)^2}. \label{SD4dim}
\end{eqnarray}
By using (\ref{CompactPropagator}) for $S\left(p_\free,p_\compact\right)$,
we generalize (\ref{SD4dim}) to
the Euclidean ladder 4-dimensional SD equation
with $d$-dimensional fermion:
\begin{eqnarray}
 m(p_\free^2) &=& \frac{3g^2}{(2\pi)^4}
  \int_{0}^{\Lambda^2} d^dk_\free \;
  \int_{0}^{\infty} d^{4-d}k_\compact \; \nonumber\\
 & &
  \frac{m(k_\free^2)}{k_\free^2 + m^2(k_\free^2)} \;
  e^{-l^2 k_\compact^2} \;
  \frac{1}{(k-p_\free)^2},
\end{eqnarray}
where
\begin{eqnarray}
 k &=& k_\free + k_\compact, \nonumber \\
 k_\free &\bot& k_\compact,  \nonumber \\
 p_\free &\bot& k_\compact,  \nonumber \\
 k_\free \cdot p_\free &=& |k_\free| |p_\free| \cos\theta.
\end{eqnarray}
This $\theta$ means the angle between $k_\free$ and $p_\free$.

We can decompose these volume measures to radius and angular components:
\begin{eqnarray}
 m(p_\free^2) &=&
  \frac{1}{4 \sqrt{\pi}} \frac{\alpha}{\alpha_c} \;
  \frac{1}{\Gamma\left(\frac{d-1}{2}\right)
           \Gamma\left(\frac{4-d}{2}\right)} \; \nonumber \\
  & & \int_{0}^{\Lambda^2} dk_\free^2 \;
      \int_{0}^{\infty} dk_\compact^2 \;
      (k_\free^2)^{\frac{d-2}{2}} \:
      (k_\compact^2)^{\frac{2-d}{2}} \;
      \frac{m(k_\free^2)}{k_\free^2 + m^2(k_\free^2)} \;
      e^{-l^2 k_\compact^2} \nonumber \\
  & & \int_{0}^{\pi} \: d\theta \; \sin^{d-2}\theta \;
      \frac{1}{k_\free^2 + k_\compact^2 + p_\free^2
           - 2|k_\free||p_\free| \cos \theta},
\end{eqnarray}
where we have defined:
\begin{eqnarray}
 \frac{3g^2}{(2\pi)^4} &=&
 \frac{3\alpha}{4\pi^3} \;=\;
 \frac{1}{4\pi^2}\frac{\alpha}{\alpha_c},
\end{eqnarray}
with $\alpha_c = \pi/3$ being the critical coupling
of ordinary 4-dimensional QED \cite{StQED}.
We can do this angular integral as
\begin{eqnarray}
 m(p_\free^2)
   &=& \frac{1}{4} \frac{\alpha}{\alpha_c} \;
       \frac{1}{\Gamma\left(\frac{d}{2}\right)
           \Gamma\left(\frac{4-d}{2}\right)} \; \nonumber \\
  & & \int_{0}^{\Lambda^2} dk_\free^2 \;
      \int_{0}^{\infty} dk_\compact^2 \;
      (k_\free^2)^{\frac{d-2}{2}} \:
      (k_\compact^2)^{\frac{2-d}{2}} \;
      \frac{m(k_\free^2)}{k_\free^2 + m^2(k_\free^2)} \;
      e^{-l^2 k_\compact^2} \nonumber \\
  & & \frac{1}{k_\free^2 + k_\compact^2 + p_\free^2} \;
      \mbox{F}\left[ \frac{1}{2}, 1, \frac{d}{2} \; ; \;
		      \frac{4 k_\free^2 p_\free^2}
		      {(k_\free^2 + k_\compact^2 + p_\free^2)^2}
	      \right], \label{SDeqQED}
\end{eqnarray}
where $F$ means the hyper-geometric function.

We approximate $m(k_\free^2)$ in the denominator of the fermion propagator
as a constant dynamical mass $m_d$
\`a la Miransky \cite{StQED}.
Then this equation is the homogeneous linear integral equation:
\begin{eqnarray}
 m(p_\free^2) &=& \frac{\alpha}{4\alpha_c}
  \int_{0}^{\Lambda^2} dk_\free^2 \;\;
  K(p_\free^2, k_\free^2) \;\; m(k_\free^2).
 \end{eqnarray}
In this equation, kernel $K$ is given by
\begin{eqnarray}
 K(p_\free^2, k_\free^2)
  &=& \frac{1}{\Gamma\left(\frac{d}{2}\right)
           \Gamma\left(\frac{4-d}{2}\right)} \;
      (k_\free^2)^{\frac{d-2}{2}} \:
      \frac{m(k_\free^2)}{k_\free^2 + m_d^{\;\:2}} \;
      \tilde{K}(p_\free^2, k_\free^2), \nonumber \\
 \tilde{K}(p_\free^2, k_\free^2)
  &:=& \int_{0}^{\infty} dk_\compact^2 \;
       e^{-l^2 k_\compact^2} \;
       (k_\compact^2)^{\frac{2-d}{2}} \nonumber \\
  & &  \frac{1}{k_\free^2 + k_\compact^2 + p_\free^2} \;
       \mbox{F}\left[ \frac{1}{2}, 1, \frac{d}{2} \;\; ; \;\;
	         \frac{4 k_\free^2 p_\free^2}
		      {(k_\free^2 + k_\compact^2 + p_\free^2)^2}
	       \right], \nonumber \\
\end{eqnarray}
where we defined a symmetric sub-kernel
$\tilde{K}(p_\free^2, k_\free^2) = \tilde{K}(k_\free^2, p_\free^2)$.

Since it is very difficult to solve this integral equation,
we approximate this sub-kernel as follows.
Note that in the 4-dimensional limit ($d \rightarrow 4$),
this sub-kernel becomes simpler:
\begin{eqnarray}
 \lim_{d\rightarrow4} \tilde{K}(p_\free^2, k_\free^2)
  &=& \frac{\theta(p_\free^2-k_\free^2)}{p_\free^2} \;+\;
      \frac{\theta(k_\free^2-p_\free^2)}{k_\free^2},
\end{eqnarray}
corresponding to
the ordinary 4-dimensional SD equation of \QEDf.
Then we approximate this sub-kernel as
\begin{eqnarray}
 \tilde{K}(p_\free^2, k_\free^2)
  &=& \theta(p_\free^2-k_\free^2) (p_\free^2)^{\frac{2-d}{2}} \;+\;
      \theta(k_\free^2-p_\free^2) (k_\free^2)^{\frac{2-d}{2}}.
  \label{KernelApprox}
\end{eqnarray}
By numerical calculation,
we can check that this approximation is good for $2 < d < 4$,
while it is exact for $d=4$.
We note that this approximated sub-kernel has
no dependence on the compactified scale $l$,
and hence the cut-off $\Lambda$ is the only scale parameter.
Thus this approximation is useful to get a phase diagram,
although it may not for getting a concrete value of the dynamical mass.

By this approximation, we can translate this integral equation to
hyper-geometric differential equation:
\begin{eqnarray}
 \frac{d}{dp_\free^2}
  \left( (p_\free^2)^{\frac{d}{2}}
   \frac{d}{dp_\free^2} m\left(p_\free^2\right)
  \right) \;+\;
  \frac{\alpha}{4\alpha_c} \frac{d-2}{2}
  \frac{ (p_\free^2)^{\frac{d}{2}} }{p_\free^2 + m_d^{\;\:2}}
  m\left(p_\free^2\right) &=& 0
\end{eqnarray}
for $(0 \leq p_\free^2 \leq \Lambda^2)$,
with an IR boundary condition:
\begin{eqnarray}
 \left[ (p_\free^2)^{\frac{d}{2}} \;
 \frac{d}{dp_\free^2}
  m\left(p_\free^2\right) \right]_{p_\free^2=0} &=& 0,
\end{eqnarray}
and a UV boundary condition at cut-off $\Lambda^2$:
\begin{eqnarray}
 \left[
 \left(
  p_\free^2 \frac{d}{dp_\free^2} \;+\;
  \frac{d-2}{2}\right) m\left(p_\free^2\right)
 \right]_{p_\free^2 = \Lambda^2} &=& 0.
\end{eqnarray}

\begin{figure}[htb] 
\begin{center}%
 \ \epsfbox{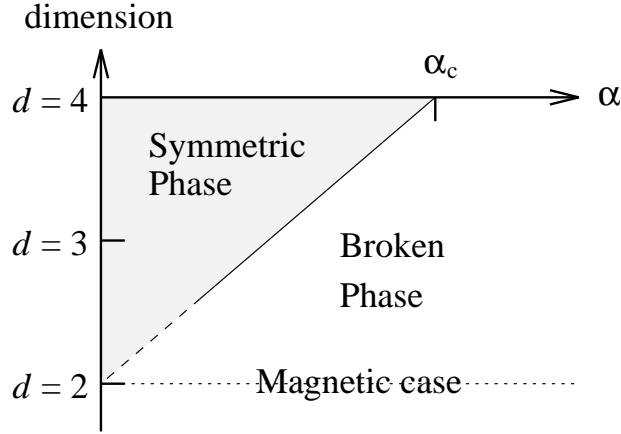}%
 \caption{Phase diagram of the chiral symmetry breaking
          in \QEDf\ with $d$-dimensional fermion.
	  This is the diagram of
          QED coupling $\alpha$ versus free fermion dimension $d$.
          $\alpha_c = \frac{\pi}{3}$ means the critical coupling of \QEDf.
          The $d=4$ line corresponds to the strong \QEDf\ phase diagram.
          The $d=2$ line corresponds to one of the \QEDf\ 
	  with the external constant magnetic field.}%
          \label{QEDphase.eps}%
\end{center}%
\end{figure}

We get the solution satisfying the IR condition:
\begin{eqnarray}
 m\left(p_\free^2\right) &=& C \;
  F\left[
     a_+, \; a_-, \; \frac{d}{2} \;\;;\;\; -\frac{p_\free^2}{m_d^{\;\:2}}
   \right], \nonumber\\
 a_{\pm} &=& \frac{d-2}{4}
  \left( 1 \pm \sqrt{1 - \frac{\alpha}{\alpha_c} \frac{2}{d-2}} \;
  \right),
\end{eqnarray}
where $C$ denotes a constant with a mass dimension containing $\Lambda$.
(The exact solution would have $C$ also containing $l$.)
This solution oscillates, when $a_{\pm}$ is imaginary, for
\begin{eqnarray}
 \frac{d-2}{2} &<& \frac{\alpha}{\alpha_c}.
\end{eqnarray}
For $d=4$, it is well known \cite{StQED} that
the non-trivial solutions of SD equation
exist only for the oscillating case.
Then we can draw the phase diagram
of dynamical breaking of the chiral symmetry (see \fig{QEDphase.eps}).

There is a linear critical line,
which is very similar to
that of the $d$-dimensional $(2<d<4)$ Gross-Neveu model
or Nambu-Jona-Lasinio model,
namely the models with naive dimensional reduction
of space-time \cite{GNlinear}.

For $d=4$ and $d=2$,
our result is consistent with the known results \cite{StQED} \cite{Miransky}.


\section{Conclusions and Discussions}

We have studied the $4$-dimensional Gross-Neveu (GN) model \cite{GN}
in the $(4-d)$-dimensional fractal space.
We adopted the $4$-dimensional type II randomized $N$-Cantor space
(\fig{Cantors.eps}-(c))
as a fractal space-time.
The reason why we used the Cantor space is that
this space is useful to control its fractal dimension.
Its fractal dimension is $4-d_\Deficit$.
Cantor spaces have holes as space-time deficit.
To consider fermions in the Cantor space,
we treated fermions in the background cut-off level potential,
namely, Cantor potential to exclude fermions from the holes.
We adopted the randomized Cantor space, 
and then we were able to treat Cantor potential as statistical objects.
There are two formulations of randomization of the Cantor space or potential;
the type I randomization and the type II one.
In the type I randomization,
we randomized the only position of holes
while preserving the number of holes.
In Appendix,
we discussed the $2$-point function of the type I randomization,
the effective action of the GN model in this randomized Cantor space and
its difficulty in the $1/N_F$ leading order.
In the type II randomization,
we put holes at random
while preserving the average number of holes.
We adopted the type II randomization
because of simplicity of the $2$-point function.

We defined $n$-point functions and
we calculated 1- and 2-point functions for this potential exactly,
where the 2-point function depends on the space-time configuration $(x,y)$
only through the integer-distance $g(x,y)$ (\ref{Distance}).
(the result of the type II is
(\ref{1ptFunction})(\ref{DefOfL})(\ref{2ptFunction})
and one of the type I is (\ref{2ptFunctionApp})).
By the local correlation approximation (\ref{LCA}),
we obtain the effective action (\ref{effectiveAction})
of the GN model in $(4-d_\Deficit)$-dimensional fractal space.
In our effective action,
the contribution of the fractal effect
appears as a shift of the $4$-fermion coupling;
when the fractal dimension becomes smaller (than $4$-dimension),
the 4-fermion coupling gets bigger.

As a result of the $1/N$ leading SD equation \cite{GN},
we got the phase diagram on the bare $4$-fermion coupling versus
fractal dimension of the space-time (\fig{fracPhase.eps}).
In this phase diagram the critical line (\ref{CriticalLineEq})
depends on the system size through the ratio $\Lambda/\lambda$.
This phase diagram implies that
the 4-fermion coupling gets bigger
as the fractal dimension becomes smaller than $4$-dimension.
There exists a critical dimension $d^* = 4-\frac{2\pi}{c}(\lambda/\Lambda)^2$,
which depends on the system size.
When we fix UV cut-off $\Lambda$ and fractal dimension,
there is a kind of critical scale $\lambda^*$ (\ref{CriticalScale})
such that in high energy the system is in the symmetric phase,
while in low energy it is in the broken phase.
In the equation of the critical line (\ref{CriticalLineEq})
when $d \leq 2$, the scale dependence is inverted.
Our analysis is valid only for small deficit dimension,
although we are also interested in whether or not
some critical phenomena happens in $d=2$.
At this moment we do not know how to treat large deficit dimension.

There is some similarity between this analysis
and the GN model in $d$ dimensions
or with naive dimensional reduction \cite{GNlinear}.
But our phase diagram depends on the system size
in contrast to the naive dimensional reduction.
Hence the naive dimensional reduction analysis does not imply
the fractal space analysis.

In the last section,
we have taken another approach to \QEDf\
with the fermions living
in non-integer dimensional space-time.
Based on the analysis of ref. \cite{Miransky},
we studied \QEDf\ with $d$-dimensional fermion
and ordinary $4$-dimensional photon.
We treated fermion which has propagator with $d$-dimensional free direction
and $(4-d)$-dimensional compactified direction (\ref{CompactPropagator}).
We have solved
the ladder SD equation (\ref{SDeqQED}).
Under a certain approximation to the kernel of the integral equation
(\ref{KernelApprox}),
we got the phase structure (see \fig{QEDphase.eps}).
This phase diagram is consistent with
the $4$-dimensional ladder QED \cite{StQED}
and the same model in the external magnetic field
with fermion dimensional reduction by 2 \cite{Miransky}.
Our phase diagram is similar to that of the
$d$-dimensional GN model
with naive dimensional reduction \cite{GNlinear}.

We think we can also apply the randomized Cantor space analysis to \QEDf\ 
by the analysis of gauged NJL \cite{GaugedNJL}.


\begin{center}
 {\Large\bf  Acknowledgments}
\end{center}
I would like to thank Prof. K. Yamawaki
for helpful suggestions and discussions,
and also for careful reading the manuscript.
I also appreciate helpful suggestions
of H. Katou and T. Ito.
I am indebted to the Japan Society for
the Promotion of Science (JSPS) for its financial support
as a JSPS Junior Fellow.
The work is supported in part
by a Grant-in-Aid for Scientific Research
from the Ministry of Education, Science and Culture.

\newpage


\appendix
\section{The Type I Randomization of Cantor Space}

In this appendix,
we discuss 
the $D$-dimensional type I randomized $N$-Cantor potential.

The type I randomization is to randomize the only position of holes
and fix the number of holes to one per parent-square.
In the type II randomization,
the number of holes per parent-square is not constant,
and only the average of number is equal to $1$.
We show the $2$-dimensional $3$-Cantor spaces;
the ordinary space, the type I and the type II randomized space
in \fig{Cantors.eps}.

The 1-point function of both randomization is the same:
\begin{eqnarray}
 \frac{1}{\Lambda} \left< V(x) \right>_{N(d)} &=&
	    \frac{1}{N^D}
	  + \frac{1}{N^D} \left( \frac{N^D-1}{N^D} \right)
	  + \cdots
          + \frac{1}{N^D} \left( \frac{N^D-1}{N^D} \right)^{G-1}
 \nonumber\\
 &=& 1 - \left( 1 - \frac{1}{N^D} \right)^G. \label{1ptFunctionAp}
\end{eqnarray}
We can also get this result by renormalization transformation:
\begin{eqnarray}
 L(0;G+1) &=& \frac{1}{N^D} \;+\; \frac{N^D-1}{N^D} L(0;G) \nonumber \\
        &=& \Reno\left[ L(0;G) \right],
\end{eqnarray}
where we defined the renormalization transformation
\begin{eqnarray}
 \Reno\left[ X \right] &:=& \frac{1}{N^D} \;+\; \frac{N^D-1}{N^D} X
\end{eqnarray}
like (\ref{Reno}).
The result of $n$-times of these transformations is
\begin{eqnarray}
 \Reno^{(n)}\left[ X \right] &:=&
   L(0;n) \;+\; \left( \frac{N^D-1}{N^D} \right)^n X.
\end{eqnarray}

The 2-point function is slightly different from type II randomization,
but the outline and the top level are the same as the type II.
The 2-point function depends only on the integer-distance $g(x,y)$
as a space-time dependence:
\begin{eqnarray}
 \frac{1}{\Lambda^2} \: \left<\: V(x)V(y) \:\right>_{N,G}
  &=& L\left( g(x,y) ; G \right)_{N}, \label{DefOfLApp}
\end{eqnarray}
where $L(g;G)$ is an increasing function of $g$.
This nature is the same as the type II.
The top level $L(G;G)$ is a self-correlation, then
\begin{eqnarray}
 L(G;G)_{N,G}
      &=&  \frac{1}{\Lambda^2} \left<V(x)V(x)\right>
     \;=\; \frac{1}{\Lambda}   \left<V(x)\right> \nonumber\\
      &=& 1 \:-\: \left(1-\frac{1}{N^D}\right)^G.
\end{eqnarray}
The top level is the same as that of the type II.

The difference from type II appears in the base level and intermediate levels.
For the base level $L(G;G)$, in the way similar to 1-point function
we have
\begin{eqnarray}
 L(1;1) &=& 0, \nonumber\\
 L(2;2) &=& \frac{2}{N^D} L(0;1)
          \;+\; \frac{N^D-2}{N^D} \left[L(0;1)\right]^2 , \nonumber\\
        &\vdots& \nonumber\\
 L(G+1;G+1) &=& \frac{2}{N^D} L(0;G)
             \;+\; \frac{N^D-2}{N^D} \left[L(0;G)\right]^2 ,
\end{eqnarray}
where $2/N^D$ means a probability of $x \in {\rm hole}$ or $y \in {\rm hole}$,
and this hole means the biggest one in any generation.
Then we get the result:
\begin{eqnarray}
 L(G;G) &=& 1 \;-\; 2 \left(\frac{N^D-1}{N^D}\right)^G
	      \;+\;   \left(\frac{N^D-2}{N^D}\right)
		  \left(\frac{N^D-1}{N^D}\right)^{2(G-1)}.
\end{eqnarray}

There is a renormalization relation among
the intermediate levels $L(g;G)$ for $0 \leq g \leq G$:
\begin{eqnarray}
 L(g;G+1) &=& \Reno \left[ L(g;G) \right],
\end{eqnarray}
where we remember the ``Reno'' transformation creates a larger structure 
preserving the fine structures.
This relation is the same as the type II.
Then the intermediate levels $L(g;G) \quad (0 \leq g < G)$
are obtained from the base level $L(g;g)$
by $(G-g)$-times $\Reno$ transformation:
\begin{eqnarray}
 L(g;G) &=& \Reno^{(G-g)} \left[ L(g;g) \right] \nonumber\\
        &=& 1 - 2 \left(\frac{N^D-1}{N^D}\right)^G
	      +   \left(\frac{N^D-2}{N^D}\right)
		  \left(\frac{N^D-1}{N^D}\right)^{G+g-2}.
\end{eqnarray}

Finally, we have resultant levels:
\begin{eqnarray}
 L(g;G) &=& \left\{	\begin{array}{ll}
           1 - 2 \left(\frac{N^D-1}{N^D}\right)^G
	     +   \left(\frac{N^D-2}{N^D}\right)
		 \left(\frac{N^D-1}{N^D}\right)^{G+g-2}
		 	& (0 < g \leq G), \\[3mm]
	   1 - \left(\frac{N^D-1}{N^D}\right)^G
			& (g = 0),
	\end{array} \right. \label{2ptFunctionApp}
\end{eqnarray}
(see \fig{2ptFunc1.eps}).

We note $L(G;G) < L(0;G)^2$, with the difference between them being
\begin{eqnarray}
 \Delta &:=& L(G;G) - L(0;G)^2 \nonumber\\
 &=& - \frac{1}{N^{2D}} \left(\frac{N^D-1}{N^D}\right)^{2(G-1)}.
\end{eqnarray}
This is the most important difference from the type II randomization.

When we shift the potential
\begin{eqnarray}
 V(x) &\longrightarrow& \tilde{V}(x) = V(x) - \left<V(x)\right>
\end{eqnarray}
for $\left<\tilde{V}(x)\right> = 0$,
the 2-point function becomes
\begin{eqnarray}
 \left<\tilde{V}(x)\tilde{V}(y)\right>
  &=& \left<V(x)V(y)\right> - \left<V(x)\right>^2.
\end{eqnarray}

Since the base level of $\left<V(x)V(y)\right>$ is close to
$\left<V(x)\right>^2$,
$\left<\tilde{V}(x)\tilde{V}(y)\right>$
has small negative base level $\Delta$
and has only one peak in the $x=y$ point.

By the ``Local Correlation Approximation'',
2-point function becomes
\begin{eqnarray}
 \left<\tilde{V}(x)\tilde{V}(y)\right>
  &\simeq& \frac{\Lambda^2}{\lambda^D} \; \left[ \;
	    \lambda^D \Delta \;+\; \Omega \: \delta^{(D)}(x-y)
	    \;\right], \label{LCAApp}
\end{eqnarray}
where $\lambda$ has a mass dimension which is a mass scale of the system size,
and $\Omega$ means dimensionless volume of this peak.

We rewrite 2-point function to get $\Omega$:
\begin{eqnarray}
\left<\tilde{V}(x)\tilde{V}(y)\right>
  &=& \Delta^G_{g(x,y)} \:+\: \Delta^G_{g(x,y)+1} \:+\: \cdots \:+\:
      \Delta^G_{G-1}    \:+\: \Delta ,
\end{eqnarray}
where
\begin{eqnarray}
 \Delta^G_g &:=& L(g;G) - L(g+1;G) \nonumber\\
 &=& \left\{
      \begin{array}{lcl}
         \frac{1}{N^D}
	  \left( \frac{N^D-2}{N^D} \right)
	  \left( \frac{N^D-1}{N^D} \right)^{G+g-2}
		& & (1 \leq g < G), \\[3mm]
         \frac{1}{N^D}
	  \left( \frac{N^D-1}{N^D} \right)^{G-1}
		& & (g = 0).
      \end{array}\right.
\end{eqnarray}
Then the volume of peak is
\begin{eqnarray}
 \Omega  &=&  \sum_{g=0}^{G-1} \; \frac{N^{Dg}}{N^{DG}} \: \Delta^G_g
   \;=\; \frac{1}{N^{2D}} \; \left( \frac{N^D-1}{N^D} \right)^{2(G-1)}.
\end{eqnarray}
This result is very similar to the one of the type II randomization.
But note that $\Omega = -\Delta$,
then the total volume of 2-point function $\Omega + \Delta$ equals to zero.

The effective action of the $4$-dimensional GN model
in this Cantor potential becomes
\begin{eqnarray}
  S_\eff[\psi]
  &:=&  \int d^4x \;\; \bar\psi i\slashDel \psi \nonumber\\
  &+&   \int d^4x \; \int d^4y \;
             \frac{1}{2N_F}
             \left[ \:
	        \left(G_\Bare + \frac{\Lambda^2}{\lambda^4} \Omega \right) \:
		  \delta(x-y) \;-\;
		\Lambda^2 \Omega
	     \right] \nonumber\\
  & &   \qquad \left(\bar\psi\psi\right)(x) \;
	       \left(\bar\psi\psi\right)(y).
  \label{effectiveAction2}
\end{eqnarray}
Then the $4$-fermion coupling in momentum space becomes
\begin{eqnarray}
 G(k) &=& \left( G_\Bare \;+\; \frac{\Lambda^2}{\lambda^4} \Omega \right)
          \;-\; (2\pi)^4 \Lambda^2 \Omega \: \tilde{\delta}(k).
\end{eqnarray}
Because of a finite space-time volume
characterized by the mass scale $\lambda$,
this delta function $\tilde{\delta(k)}$ at $k=0$ stands for a finite value:
\begin{eqnarray}
(2\pi)^4 \tilde{\delta}(k)
 &=& \int dx \; e^{i\:kx} 
 \;=\; \left\{
       \begin{array}{cl}
         0           & (k \ne 0), \\
         1/\lambda^4 & (k  =  0).
       \end{array} \right.
\end{eqnarray}
The $1/N_F$ leading SD equation now reads
\begin{eqnarray}
 i S(p) - \pSlash &=&
\setlength{\unitlength}{0.0125in}%
\begin{picture}(200,50)(0, -2)
\thicklines
\put(  0,  0){\line( 1, 0){ 80}}
\put( 40, 15){\circle{30}}
\put( 40,  0){\circle*{3.5}}
\put( 40, -4){\makebox(0,0)[ct]
              {$\scriptstyle G_\Bare + \Lambda^2/\lambda^4 \; \Omega $}}
\put(100,  0){\makebox(0,0)[c]{$+$}}
\put(120,  0){\line( 1, 0){ 80}}
\put(160, 35){\circle{30}}
\put(160, 20){\circle*{3.5}}
\put(160,  0){\circle*{3.5}}
\multiput(160, 20)(0.00000,-3.00000){8}{\circle*{1}}
\put(165, 10){\makebox(0,0)[l]
              {$\scriptstyle \tilde{\delta}(0) $}}
\put(160, -4){\makebox(0,0)[ct]
              {$\scriptstyle -(2\pi)^4 \Lambda^2 \Omega $}}
\end{picture} \nonumber \\[5mm]
  &=& G_\Bare \:
      \int \frac{d^4k}{(2\pi)^4} \;
      {\rm tr} \: S(k),	\label{NonLocalSDeq}
\end{eqnarray}
where the solid lines mean fermion line which takes the form
\begin{eqnarray}
 S^{-1}(p) &=& i \: \left( B(p^2) \:-\: A(p^2) \pSlash \right),
\end{eqnarray}
and the dotted lines mean the constant non-local 4-fermion interaction.
Then the SD equation (\ref{NonLocalSDeq}) is the same as
the one of the ordinary $4$-dimensional GN model.

In the type I randomized Cantor space,
we conclude that
the $1/N_F$ leading SD equation cannot describe the dependence on
the fractal dimension.
We must analyze next to the leading order of $1/N_F$ expansion.


\newpage



\end{document}